\documentclass[a4paper,10pt]{article}
\usepackage{psfrag}
\usepackage{graphicx}
\usepackage{graphics}
\usepackage{bm}
\usepackage{color}
\usepackage{amssymb}
\usepackage{amsmath}
\usepackage{float}

\usepackage[T1]{fontenc}
\usepackage[latin1]{inputenc}
\usepackage{amsthm}
\usepackage{braket}
\usepackage{lmodern}
\usepackage{fancyhdr} 
\usepackage{rotating}
\newcommand{\gsim}{\raise.3ex\hbox{$>$\kern-.75em\lower1ex\hbox{$\sim$}}}
\newcommand{\lsim}{\raise.3ex\hbox{$<$\kern-.75em\lower1ex\hbox{$\sim$}}}

\usepackage{vmargin}         
\setmarginsrb{2cm}{2cm}{2cm}{2cm}{0cm}{0cm}{0cm}{1cm}

\DeclareMathOperator{\Cin}{Cin}

\begin{document}

\title{Gravitational power from cosmic string loops with many kinks}
\author{Alejandro Boh\'e\footnote{bohe@apc.univ-paris7.fr}
{\small {}}\\
{\small {\it  APC\ \footnote{Universit\'e 
Paris-Diderot, CNRS/IN2P3,  CEA/IRFU and Observatoire de Paris} ,10 rue Alice Domon et L\'eonie Duquet,
 75205 Paris Cedex 13, France}}\\
 }
\maketitle

\begin{abstract}
We investigate the effect of a large number of kinks on the gravitational power radiated by cosmic string loops. We show that the total power radiated by a loop with $N$ left-moving and right-moving kinks is proportional to $N$ and increases with the typical kink angle. We then apply these results to loops containing junctions which give rise to a proliferation of the number of sharp kinks. We show that the time of gravitational decay of these loops is smaller than previously assumed. 
In light of this we revisit  the gravitational wave burst predictions from a network containing such loops.
We find  there is no parameter regime in which the rate of individual kink bursts is enhanced with respect to standard networks. By contrast, there remains a region of parameter space for which the kink-kink bursts dominate the stochastic background. 
Finally, we discuss the order of magnitude of the typical number of sharp kinks resulting from kink proliferation on loops with junctions.

\end{abstract}

\section{Introduction}
Cosmic strings are one-dimensional topological defects of cosmic size originally introduced by Kibble \cite{Kibble:1976sj} which may have been produced during spontaneous symmetry-breaking phase transitions in the early Universe. Recently, it has been realized that fundamental objects from string theory could, in certain scenarios, expand to cosmic size and also play the role of cosmic (super-)strings (see \cite{Copeland:2009ga} and references therein). A very promising strategy to detect them relies on their gravitational wave emission, and in particular on their emission of high frequency bursts at cusps and kinks \cite{Damour:2000wa,Siemens:2006vk} which could be accessible to the advanced versions of the network of ground based interferometers LIGO-VIRGO and to the future space interferometer LISA in a large region of parameter space. Although cosmic strings have never been observed so far, a lot of effort is made in this direction since their detection would provide unique insight on either high energy physics or string theory.

Several authors have studied how certain properties of cosmic superstrings could help distinguishing their gravitational wave burst signal from that of field theory cosmic strings, such as their reduced reconnection probability \cite{Damour:2004kw}, their reconnection process itself \cite{Jackson:2009fk} or the fact that they can move in extra dimensions \cite{OCallaghan:2010fk}. Another feature of superstrings that drastically affects their gravitational wave burst signal is the existence of junctions between strings. A cosmic superstring network contains F- and D-strings as well as bound states made up of both types called (p,q) strings. The crossing of two strings of different types can lead to the creation of a new segment of string joining the original ones at two vertices also called Y-junctions \cite{Copeland:2006eh,Copeland:2006if,Copeland:2007nv,Bevis:2008hg,Salmi:2007ah}.  Note that junctions can also be formed in specific models of symmetry-breaking \cite{Saffin:2005cs,Copeland:2005cy,Rajantie:2007hp}.

The impact of junctions on the gravitational wave burst emission from a cosmic string loop was investigated in \cite{Binetruy:2009vt,Binetruy:2010bq}. It was shown that the most important effect they induce is the exponential proliferation of large amplitude kinks. In \cite{Binetruy:2010cc}, the effect of kink proliferation on the signal emitted by a whole network of loops was investigated. It was found that, for models where the fraction $q$ of loops that contain junctions and the average number $k'$ of sharp kinks on these loops satisfy the condition $qk'>1$, then the observational predictions differ from the standard case examined in \cite{Damour:2000wa,Damour:2004kw,Damour:2001bk}, mainly due to the presence of a strong stochastic background resulting from the incoherent superposition of bursts produced at kink-kink collisions potentially hiding individual bursts in certain regions of parameter space.

One of the key assumptions in this analysis was that the lifetime of the loops with junctions (or of their daughter loops in the case where the junctions unzip) was unaffected by the large number of kinks on them. To be more precise, it was assumed that the gravitational power radiated by these loops was given by the standard formula $P=\Gamma G \mu^2$ with $\Gamma\sim50$, a result first obtained by Vachaspati and Vilenkin in \cite{Vachaspati:1985} and a priori only valid for rather smooth loops. This rate of energy loss determines the lifetime of the loops in the network and therefore enters crucially in the estimation of the density of loops at any time and ultilmately in the rate of observed events.

In this paper, we investigate the validity of this assumption by computing the radiation from piecewise linear loops with a large number of sharp kinks. The radiation from kinky loops was already explored in the past, but always with a different focus. The first calculations date back to the work of Vachaspati and Garfinkle \cite{Garfinkle:1987yw} who showed that the power radiated by kinky cuspless loops was finite in every direction and computed analytically the power radiated by a particular class of loops with $4$ kinks in total. Other simple configurations were studied in \cite{Allen:1994bs} by Allen, Casper and Ottewill. In \cite{Allen:1994iq}, Allen and Casper derived an analytical expression for the total power radiated by any piecewise linear loop made of $N$ straight segments as a sum of $\mathcal{O}(N^4)$ polynomial and logarithmic terms. Their idea was to provide a way of computing the power from smooth loops by approximating them by increasingly complex polygons with angles at the vertices tending to zero, a situation completely different than the one we are considering here since we are interested in a large number of sharp kinks. Although their formula can be used to compute the power from any given loop with many sharp kinks, its complicated form does not allow to determine analytically the generic large $N$ behavior for such loops.

Here, we develop a completely different approach, based on an analytical averaging of the power over loop trajectories with $N$ sharp kinks leading to an expression where the large $N$ behavior can be extracted. This is the first time that the power radiated by loops with many sharp kinks is investigated using a fully relativistic formalism, namely Weinberg's formula and our main result is that the power radiated is proportional to the number of kinks\footnote{We note that the linear dependence of the power on the number of kinks had already been proposed in \cite{Quashnock:1990qy} but the argument was based on the quadrupole radiation formula which only applies to slowly moving sources and whose validity is therefore not clear in the context of cosmic strings.}.

We then apply our results to loops with junctions and revisit the estimate for their number density and the gravitational wave signal predictions made in \cite{Binetruy:2010cc}. In particular, we investigate the modifications to the rate of individual kink bursts and the amplitude of the stochastic backgrounds formed by the superposition of kink bursts and of kink-kink bursts. 

Finally, our results provide a new step towards understanding how long kink proliferation can last and towards estimating the resulting value of the parameter $k'$. Indeed, the expression of the power radiated as a function of the number of kinks enables us to compute the energy lost as kinks proliferate on the loop and therefore to estimate the time it takes for a loop to decay by gravitational radiation. We also discuss qualitatively how other mechanisms such as intercommutations, unzipping of junctions and gravitational backreaction affect our estimate.

The paper is organized as follows. In section \ref{sec:radiation}, we study the radiation from piecewise loops with a large number of kinks. In section 3, we use our results to revisit the observational predictions taking into account the shorter lifetime of the loops with junctions. In Section 4, we discuss the implications of our results on the duration of kink proliferation and therefore on the number of kinks. Section 5 contains our conclusions.

\section{Radiation by a loop with many kinks}
\label{sec:radiation}
In this section, we compute the power radiated from periodic piecewise loops with $N$ kinks. Sec. \ref{subsec:classofloops} is devoted to the definition of the class of loops that we consider. Then, in Sec. \ref{subsec:powerradiated}, we use Weinberg's formula to derive a general expression for the power radiated by the loops in this class. In Sec. \ref{subsec:analytical}, we analytically investigate the dependence of the power on the number of kinks. This section contains our main result, namely the fact that for large $N$, the power becomes proportional to $N$. In Sec. \ref{subsubsec:num}, we use numerical calculations to investigate how our result depends on the amplitude of the kinks. Finally, in Sec. \ref{subsubsec:complit}, we comment on the applicability of our results to the case of loops with junctions.

\subsection{Piecewise constant loops: the setup}
\label{subsec:classofloops}
Let us start by setting the notations and defining the class of loops that we use. We consider a cosmic string loop whose dynamics is governed by the Nambu action. We assume flat spacetime background geometry $(-+++)$ (we only consider loops small compared to the horizon size) and we use the standard conformal time gauge to parametrize the string's worldsheet so that the string is described by its spatial coordinates $\textbf{x}(\sigma,t)$ where $t$ coincides with Lorentz time. The gauge constraints can then be written (with $'$ and $ \dot{ }$ standing for derivatives with respect to  $ \sigma$ and $t$)
\begin{eqnarray}
\textbf{x}' \cdot \dot{\textbf{x} }& = & 0, \\
\textbf{x}^{'2} + \dot{\textbf{x} }^{2} & = & 1.
\end{eqnarray}
The wave-like equation of motion leads to the form
\begin{equation}
\label{ }
\textbf{x}(\sigma,t)=\frac{1}{2}\big(\textbf{x}_+(\sigma+t)+\textbf{x}_-(\sigma-t)\big)
\end{equation}
with $\textbf{x}_+^{'2}=\textbf{x}_-^{'2}=1$ in order to satisfy the gauge constraints (here $'$ denotes a derivative to the only variable of $\mathbf{x}_+$ or $\mathbf{x}_-$). 

As can easily be checked, the total power radiated will be independent of the invariant length of the string $L$ so, in order to simplify the expressions, we set $L=2 \pi$ so that at any time, $\sigma\in[0,2\pi]$. The evolution is periodic and $\mathbf{x}_+$ and $\mathbf{x}_+$ are periodic functions
\begin{eqnarray}
\label{periodicity}
\mathbf{x}_\pm(0)&=& \mathbf{x}_\pm(2\pi).
\end{eqnarray}
The definition of these two functions on the interval $[0,2\pi]$ is set by the inital conditions ($\mathbf{x}'(t=0,\sigma)$ and $\dot{\mathbf{x}}(t=0,\sigma)$). The whole evolution of the loop can then be determined trivially using periodicity. We now define the class of initial conditions that we use.\\

Let $ \mathbf{A}_{\pm,n}, n=0..N-1$ be two sets of $N$ unit vectors such that $\sum_{n=0}^{N-1} \mathbf{A}_{\pm,n}=\mathbf{0}$. We can define the derivative of the functions $ \mathbf{x}_\pm(u)$ as a piecewise constant functions:
\begin{equation}
\label{A}
\mathbf{x}_\pm'(u)= \mathbf{A}_{\pm,n} \qquad \text{ for  } u\in\left[\frac{2n\pi}{N},\frac{2(n+1)\pi}{N}\right]
\end{equation}
By integration, $ \mathbf{x}_\pm$ are piecewise linear functions given by (the integration constants vanish in the center of mass frame of the loop)
\begin{equation}
\label{a}
\mathbf{x}_\pm(u)=\frac{2\pi}{N}\sum_{q=0}^{n-1} \mathbf{A}_{\pm,q} + \left(u-\frac{2n\pi}{N}\right) \mathbf{A}_{\pm,n} \qquad \text{ for  } u\in\left[\frac{2n\pi}{N},\frac{2(n+1)\pi}{N}\right]
\end{equation}
(with the convention $\sum_{q=0}^{-1}\mathbf{A}_{\pm,q} =\mathbf{0}$). The condition \eqref{periodicity} is clearly satisfied since $\sum_{n=0}^{N-1} \mathbf{A}_{\pm,n}=\mathbf{0}$.\\

A loop defined like this has $N$ left moving and $N$ right moving kinks evenly spaced on the string.\footnote{We are mainly interested in such symmetric configurations because we expect any mechanism responsible for the creation of many kinks to be symmetric in the left and the right movers. In particular, kink proliferation on loops with junctions that we discuss in Sec. \ref{subsubsec:complit} creates similar distributions of left-moving and right-moving kinks. We will however comment on the power radiated by loops with $N\gg1$ left-moving kinks and only a small number of  right moving kinks at the end of Sec. \ref{subsec:analytical}.}

\subsection{Radiated power}
\label{subsec:powerradiated}
Because the evolution is periodic the frequencies of emission are all multiples of the fundamental mode of the string $\omega_m=m\omega_0$ where $\omega_0=\frac{4\pi}{L}=2$. The total power radiated by the loop per unit solid angle in the direction given by the unit vector $\mathbf{k}$ is a sum over all such frequencies and can be computed using Weinberg's formula \cite{Weinberg}

\begin{equation}
\label{Weinberg}
\frac{dP}{d\mathbf{k}}=\sum_{m=1}^{\infty}\frac{G \omega_0^2m^2}{\pi}\Lambda_{ij,lp}(\mathbf{k})T^{ij*}(\mathbf{k}_m,\omega_m)T^{lp}(\mathbf{k}_m,\omega_m)
\end{equation}
where $\mathbf{k_m}=\omega_m\mathbf{k}$, $T^{ij}(\mathbf{k},\omega)$ is the spacetime Fourier transform of the stress-energy tensor of the string and $\Lambda_{ij,lp}(\mathbf{k})$ is a projection operator defined as 
$\Lambda_{ij,lp}(\mathbf{k})=P_{il}P_{jp}-\frac{1}{2}P_{ij}P_{lm}$ where $P_{ij}=\delta_{ij}-k_ik_j$. Note that $P_{ij}k^{i}=0$: $P_{ij}$ is the projector on the plane perpendicular to $\mathbf{k}$.
Here $\delta$ is the Kronecker symbol and all indexes $i,j,l,p$ range from 1 to 3.\\

The (spacetime Fourier transform of the) stress-energy tensor can be conveniently written under the factorized form \cite{Damour:2000wa}
\begin{equation}
\label{Tij}
T^{ij}(\mathbf{k}_m,\omega_m)=\frac{\mu}{2\pi} I_+^{(i}(\mathbf{k},m)I_-^{j)}(\mathbf{k},m),
\end{equation}
where $^{(ij)}$ denotes symmetrization with respect to the indexes $i$ and $j$ and
\begin{equation}
I_\pm^i(\mathbf{k},m)=\int_0^{2\pi} x_\pm'^i(u)e^{\frac{i}{2}m\omega_0(\pm u-\mathbf{k}\cdot \mathbf{x}_\pm(u))}du .
\end{equation}

The integrals above can be easily computed using \eqref{A} and \eqref{a} and yield
\begin{equation}
\label{I+}
I_\pm^i=\frac{i}{m}\sum_{n=0}^{N-1} C_{\pm,n}^i e^{\frac{2i\pi m}{N}\left(\pm n-\mathbf{k}\cdot \left(\sum_{s=0}^{n-1} \mathbf{A}_{\pm,s}\right)\right)}
\end{equation}
where we defined
\begin{equation}
\label{Cn}
C_{\pm,n}^i=\frac{A_{\pm,n}^i}{\mathbf{k}\cdot \mathbf{A}_{\pm,n}\mp1}\left(e^{\frac{2i\pi m}{N}(\pm1-\mathbf{k}\cdot \mathbf{A}_{\pm,n})}-1\right).
\end{equation}
Both $I_\pm$ and the $C_{\pm,n}$  depend on $\mathbf{k}$ and $m$ but from now on, in an effort to avoid heavy notation as much as possible, we leave this dependence implicit. Note that $C_{\pm,n}^i$ is not singular at $\mathbf{k}=\pm \mathbf{A}_{\pm,n}$, and therefore $I_\pm^i$ is not singular at any $\mathbf{k}= \pm \mathbf{A}_{\pm,n}$. More precisely, if $ \mathbf{k}= \mathbf{A}_{+,n_0}$,  we have $I_+^i(\mathbf{k}= \mathbf{A}_{+,n_0},m)= \frac{2\pi}{N}A_{+,n_0}^i e^{\frac{2i\pi m}{N} \left(n_0-\mathbf{k}\cdot \left(\sum_{s=0}^{n-1} \mathbf{A}_{+,s}\right)\right)}+\mathcal{O}(\frac{1}{m})$. Therefore, although $I_+$ is finite in those directions, it doesn't tend to zero as $m$ goes to infinity. However, the series in \eqref{Weinberg} still converges because the operator $\Lambda$ projects out this first term since it is proportional to $ \mathbf{A}_{+,n_0}= \mathbf{k}$. As a consequence, $\frac{dP}{d \mathbf{k}}$ is finite in every direction in perfect agreement with the findings of \cite{Garfinkle:1987yw} in the case $N=2$ (except in the degenarate case of a persistant cusp when one of the $ \mathbf{A}_{+,n}$ is exactly equal to one of the $ \mathbf{A}_{-,q}$).\\

Finally, the \emph{total} power radiated by the loop is the integral of \eqref{Weinberg} over all directions
\begin{equation}
\label{totalpower}
P=\int \frac{dP}{d\mathbf{k}}d \mathbf{k}.
\end{equation}

\subsection{Dependence on the number of kinks: analytical study}
\label{subsec:analytical}
\subsubsection{Loop distribution}
\label{subsubsec:distrib}
We now would like to average over the class of loop trajectories defined in Section \ref{subsec:classofloops} for fixed values of the number of kinks $N$ and study how the total power radiated varies with $N$. In order to do so, we first need to specify the weight of each loop of the class or equivalently, a random procedure to construct such loops. This raises the question of the naturalness of the probability distribution that we will use: what do \emph{realistic} kinky loops look like? Are the kink angles typically large or small? For now, we do not address these complex questions (we will focus starting from Sec. \ref{subsubsec:complit} on a particular mechanism creating many kinks, namely kink proliferation on loops with junctions) and make the arbitrary choice of using a "flat" distribution (in a sense that we explain in the next paragraph) for which the average can be computed analytically. In Sec. \ref{subsubsec:num}, we use numerical computations to discuss how different distributions can affect the result.\\

The simplest idea to obtain a flat distribution is to draw randomly each of the $ \mathbf{A}_{+,n}$ and $ \mathbf{A}_{-,n}$ for $n=0..N-1$ with uniform probability on the unit sphere. However, we must satisfy the conditions $\sum_{n=0}^{N-1} \mathbf{A}_{\pm,n}=\mathbf{0}$. In an attempt to generalize the trajectories used in \cite{Garfinkle:1987yw} in the case $N=2$, we use the following construction: we restrict our attention to even values of $N$ and  draw each of the $ \mathbf{A}_{+,n}, n-0..N/2-1$ and $ \mathbf{A}_{-,n}, n-0..N/2-1$ with uniform probability on the unit sphere. For $n=N/2..N-1$, we then define the remaining $ \mathbf{A}_{\pm,n}= -\mathbf{A}_{\pm,n-N/2}$ so that the sums vanish. To be precise, this yields a flat distribution but only over a subset of our initial class. Though such a construction is probably the simplest, other are possible and the robustness of our results to the distribution of loops will be discussed in the next section.

\subsubsection{Average power}
Since there is no preferred direction in our loop distribution, we expect $\langle\frac{dP}{d \mathbf{k}}\rangle$ to be independent of $ \mathbf{k}$. We can therefore set $\mathbf{k}=(1,0,0)$ from now on to simplify our calculations. From \eqref{Weinberg} and \eqref{Tij}, we have
\begin{equation}
\left\langle\frac{dP}{d \mathbf{k}}\right\rangle=\frac{G\mu^2}{4\pi^3} \sum_{m=1}^{+\infty} m^2 \Lambda_{ij,lp} \left \langle\left(I_+^{i*}I_-^{j*}+I_+^{j*}I_-^{i*}\right)\left(I_+^lI_-^p+I_+^pI_-^l\right)\right \rangle
\end{equation}
where from now on the brackets $\langle\cdot\rangle$ denote an average over the class of loops.
Since the $ \mathbf{A}_+$ and $ \mathbf{A}_-$ are drawn independently, the averages over $I_+$ and $I_-$ factorize and we are left with
\begin{equation}
\label{pleindeI}
\left\langle\frac{dP}{d \mathbf{k}}\right\rangle=\frac{G\mu^2}{4\pi^3} \sum_{m=1}^{+\infty} m^2 \Lambda_{ij,lp} \left(
\left \langle I_+^{i*}I_+^l\right \rangle \left \langle I_-^{j*}I_-^p\right \rangle
+\left \langle I_+^{i*}I_+^p\right \rangle \left \langle I_-^{j*}I_-^l\right \rangle
+\left \langle I_+^{j*} I_+^l\right \rangle \left \langle I_-^{i*}I_-^p\right \rangle
+\left \langle I_+^{j*}I_+^p\right \rangle \left \langle I_-^{i*}I_-^l\right \rangle
 \right).
\end{equation}
Therefore, the building blocks of our calculation will be $\left \langle I_+^{i}I_+^{j*} \right \rangle$ and $\left \langle I_-^{i}I_-^{j*} \right \rangle$ which, as we explain now, are complex conjugate quantities. Indeed, by remarking that (with obvious notations) $C_{-,n}^i\left(\mathbf{A}_{-,n}\right)=\left(C_{+,n}^i\left(-\mathbf{A}_{-,n}\right)\right)^*$ and $I_-^i\left(\left\{\mathbf{A}_{-,n}\right\}\right)=-I_+^i\left(\left\{-\mathbf{A}_{-,n}\right\}\right)^*$
we can immediately conclude that
\begin{eqnarray}
\left \langle I_-^iI_-^{j*} \right \rangle&=&\left \langle I_-^i\left(\left\{\mathbf{A}_{-,n}\right\}\right) I_-^{j}\left(\left\{\mathbf{A}_{-,n}\right\}\right)^*  \right \rangle\\
 & = & \left \langle I_+^i\left(\left\{-\mathbf{A}_{-,n}\right\}\right)^* I_+^{j}\left(\left\{-\mathbf{A}_{-,n}\right\}\right)  \right \rangle\\
 & = & \left \langle I_+^i\left(\left\{-\mathbf{A}_{-,n}\right\}\right) I_+^{j}\left(\left\{-\mathbf{A}_{-,n}\right\}\right)^*  \right \rangle ^*\\
 & = & \left \langle I_+^iI_+^{j*} \right \rangle^*
\end{eqnarray} 
where in going from the third to the last line we used the fact that any set of vectors $\left\{\mathbf{A}_{-,n}\right\}$ and the set of their opposites $\left\{-\mathbf{A}_{-,n}\right\}$ has the same probability with our procedure.\\

We will now focus on one of these objects,  say $\left \langle I_+^iI_+^{j*} \right \rangle$, and drop all the indices $+$ so that we now use the notation  $C_n^i$ instead of $C_{+,n}^i$ and $\mathbf{A}_n$ instead of $\mathbf{A}_{+,n}$. Because the calculation is somewhat long and technical, we save all the details for Appendix \ref{sec:appendixA} and simply present the outline of the calculation here before jumping to the result. Starting from formula \eqref{I+}, it is clear that $\left \langle I_+^iI_+^{j*} \right \rangle$ will be a sum of average values of products of terms of the form $C_n^iC_{n'}^{j*} $ with several exponentials of scalar products of the type $\mathbf{k}\cdot \mathbf{A}_s$. The first step is to factorize these average values as much as possible using the fact that the random unit vectors $\mathbf{A}_n$ for $n$ between $0$ and $N/2-1$ are drawn independently
to obtain a product of expectation values each involving only one of the vectors $\mathbf{A}_n$. We can then use the symetries to reduce the number of components that we need to calculate. Indeed, since we had set $\mathbf{k}=(1,0,0)$ and we will eventually contract $ \left \langle I_+^iI_+^{j*} \right \rangle$ with the operator $\Lambda_{ij,lp}$ which projects away any term proportional to $\mathbf{k}$, any component of $\left \langle I_+^iI_+^{j*} \right \rangle$ bearing an index $1$ will disappear. Furthermore, after remarking that directions $2$ and $3$ are equivalent, it becomes clear that the only relevant quantity to compute is $\left \langle I_+^2I_+^{2*} \right \rangle=\left \langle I_+^3I_+^{3*} \right \rangle$ which can be expressed in terms of the three following functions computed in Appendix \ref{appendix:smusigmaasfmN}
\begin{equation}
s=\left \langle e^{\frac{2i\pi m}{N}\mathbf{k}\cdot\mathbf{A}_n}\right \rangle= \frac{N}{2\pi m} \sin\left(\frac{2\pi m}{N}\right),
\end{equation}
\begin{equation}
\mu= \left \langle C_n^{2}C_{n+N/2}^{2*} e^{\frac{2i\pi m}{N}\mathbf{k}\cdot\mathbf{A}_n} \right \rangle=\cos\left(\frac{2\pi m}{N}\right)-\frac{N}{2\pi m}\sin\left(\frac{2\pi m}{N}\right),
\end{equation}
and finally
\begin{equation}
\sigma =\left \langle C_n^2 C_{n}^{2*}\right \rangle=\Cin\left(\frac{4\pi m}{N}\right) +\frac{N}{4\pi m}\sin\left(\frac{4\pi m}{N}\right) -1\, ,
\end{equation}
where the special cosine integral is defined as
\begin{equation}
\Cin(x)=\int_0^x\frac{1-\cos t}{t}dt.
\end{equation}
Given our procedure to draw the vectors $ \mathbf{A}_n$, it is clear that none of the average values defined here actually depends on the index $n$. Also note that all three quantities $\sigma, \mu$ and $s$ are real. 

Now jumping to the result of the calculation, we have
\begin{equation}
\label{IIQ}
\left \langle I_+^iI_+^{j*} \right \rangle=  \frac{N}{m^2} \left(\sigma+(-1)^m s^{N/2-1} \mu \right) \delta^{ij}(1-\delta^{i1})+ Q^{ij}
\end{equation}
where the tensor $Q^{ij}$ will be projected away by $\Lambda_{ijlp}$. In other words, we can now replace in our calculations $\left \langle I_+^iI_+^{j*} \right \rangle$ by the diagonal tensor
\begin{equation}
\label{diagonalII}
\frac{N}{m^2} \left(\sigma+(-1)^m s^{N/2-1} \mu \right) \left(\begin{array}{ccc}0 & 0 & 0 \\0 & 1 & 0 \\0 & 0 & 1\end{array}\right).
\end{equation}
Note in particular that because the part of $\left \langle I_+^iI_+^{j*} \right \rangle$ that is not projected away is real, we can in practice also replace $\left \langle I_-^iI_-^{j*} \right \rangle$ in \eqref{pleindeI} by the expression given in \eqref{diagonalII}.\\

When performing the contraction in \eqref{pleindeI}, the only term that appears is therefore the square of the factor in \eqref{diagonalII} and it appears $8$ times. This leaves us with 

\begin{equation}
\label{contracte}
\left\langle\frac{dP}{d \mathbf{k}}\right\rangle=\frac{2G\mu^2}{\pi^3} \sum_{m=1}^{+\infty} \frac{N^2}{m^2}\left[\sigma(m,N) + (-1)^m s(m,N)^{N/2-1} \mu(m,N) \right]^2
\end{equation}
where we now indicated explicitely the dependence on $m$ and $N$. Note that all three quantities $\sigma$, $\mu$ and $s$ are only functions of $\frac{m}{N}$ so from now on we will slightly abuse notations and write $\sigma(m/N)$ for example. 

\subsubsection{Large $N$ behavior}
Our goal is now to understand the large $N$ behavior of \eqref{contracte}, for which it is convenient to expand the square of the brackets, thus expressing $\left\langle\frac{dP}{d \mathbf{k}}\right\rangle$ as the sum of three terms
\begin{equation}
\label{contractethreeterms}
\left\langle\frac{dP}{d \mathbf{k}}\right\rangle=\frac{2G\mu^2}{\pi^3}\left( \sum_{m=1}^{+\infty} \frac{N^2}{m^2}\sigma^2\left(\frac{m}{N}\right) +\sum_{m=1}^{+\infty} \frac{N^2}{m^2}\mu^2\left(\frac{m}{N}\right) s\left(\frac{m}{N}\right)^{N-2}  +  2\sum_{m=1}^{+\infty} (-1)^m \frac{N^2}{m^2}\sigma\left(\frac{m}{N}\right) \mu\left(\frac{m}{N}\right) s\left(\frac{m}{N}\right)^{N/2-1} \right)
\end{equation}

We will now show that the dominant term in the $N\rightarrow\infty$ limit is the first one. The behavior of this term is simple to obtain because the dependence on $m$ and $N$ is entirely of the form $m/N$ so we can use a Riemann integral to show that it is proportional to $N$. For the other terms, precisely determining an equivalent is much more complicated but  we just need to show that they are subdominant. For the sake of simplicity, we chose to present here only qualitative arguments for this instead of a complete but much more technical proof.\\

Let us start with the first sum. In the large $N$ limit, we clearly have
\begin{equation}
\label{dominantintegral}
\sum_{m=1}^{+\infty} \frac{N^2}{m^2}\sigma^2\left(\frac{m}{N}\right) \approx N \int_0^{\infty} f(u)du
\end{equation}
with
\begin{equation}
f(x)=\frac{\sigma^2(x)}{x^2}=\frac{1}{x^2}\left[\Cin(4\pi x) +\frac{1}{4\pi x}\sin(4\pi x) -1\right]^2
\end{equation}
It is straightforward to check that $f(x)=\mathcal{O}_{0}(x^2)$ and $f(x)=\mathcal{O}_{\infty}\left(\frac{\ln(x)^2}{x^2}\right)$ so the integral in \eqref{dominantintegral} is perfectly well defined. 

For the second sum, the power $N-2$ prevents us from using a similar technique.
However, we can understand qualitatively why this term is subdominant. The factor $s\left(\frac{m}{N}\right)^{N-2}$ quickly goes to zero at large $N$ because $s(x)=\sin(2\pi x)/(2\pi x)\leq1$ except when $m/N\ll1$. Therefore, only the modes $m\ll N$ contribue to the sum but as $N$ grows, the number of such modes also grows. However, becaues in the limit $x\rightarrow 0$ the function $\mu^2(x)/x^2$ goes to zero, their contribution tends to zero and therefore the total sum goes as $N$ times a term of limit zero. A similar argument shows that the third term is also subdominant.\\

As a conclusion, in the large $N$ limit, we have
\begin{equation}
\label{dPdominantN}
\left\langle\frac{dP}{d \mathbf{k}}\right\rangle\approx G\mu^2  \left(\frac{2}{\pi^3}\int_0^{\infty} f(u)du\right) N
\end{equation}
The integration over $\mathbf{k}$ is trivial since $\left\langle\frac{dP}{d \mathbf{k}}\right\rangle$ is independent of $\mathbf{k}$ and yields
\begin{equation}
\label{PdominantN}
\left\langle P \right\rangle \approx G\mu^2 \left(\frac{8}{\pi^2} \int_0^{\infty} f(u)du \right) N 
\end{equation}\\

The integral can be computed numerically to obtain our final result for the power radiated by loops with $N\gg1$ left-moving and right-moving kinks\footnote{If we consider loops with $N$ left-moving kinks but only $2$ right-moving kinks, a similar computation leads to $\left\langle\frac{dP}{d \mathbf{k}}\right\rangle=\frac{2G\mu^2}{\pi^3} \sum_{m=1}^{+\infty} \frac{N}{m^2}\left[\sigma+ (-1)^m s^{N/2-1} \mu \right]\Cin(2\pi m)$ instead of Eq. \eqref{contracte} which can be shown to grow only logarithmically with $N$. This is consistent with the results of \cite{Siemens:2001dx} that small scale structure only traveling in one direction on the string does not lead to efficient energy loss.}

\begin{equation}
\label{PdominantNnum}
\left\langle P \right\rangle \approx G\mu^2 \Gamma'\, N \qquad \text{ with } \qquad \Gamma'\approx 13
\end{equation}

\subsection{Numerical study}
\label{subsubsec:num}
\subsubsection{Computation for a given loop}
We now explain how the power radiated by a given loop belonging to the class defined in Section \ref{subsec:classofloops} can be computed numerically. This calculation was performed analytically by Garfinkle and Vachaspati in \cite{Garfinkle:1987yw} in the case $N=2$. In this section, we generalize their method to an arbitrary value of $N$. The infinite sum over $m$ in \eqref{Weinberg} can still be performed analytically. The integral over $\mathbf{k}$ is done numerically over a grid.

Let us start by rewriting \eqref{I+} in the following way

\begin{equation}
I_\pm^i=\frac{i}{m}\sum_{n=0}^{N-1} D_{\pm,n}^i \left(e^{i m \alpha_{\pm,n}}-e^{i m \widetilde{\alpha}_{\pm,n}}\right)
\end{equation}
where we defined 
\begin{equation}
\label{alphan}
\alpha_{n}^\pm=\frac{2\pi}{N}\left[\pm(n+1)-\mathbf{k}\cdot \mathbf{A}_{\pm,n}-\mathbf{k}\cdot \left(\sum_{s=0}^{n-1} \mathbf{A}_{\pm,s}\right)\right] \qquad
 \text{and} \qquad \widetilde{\alpha}_{n}^\pm=\frac{2\pi}{N}\left[\pm n-\mathbf{k}\cdot \left(\sum_{s=0}^{n-1} \mathbf{A}_{\pm,s}\right)\right]
\end{equation}
as well as
\begin{equation}
\label{Dn}
D_{\pm,n}^i=\frac{A_{\pm,n}^i}{\mathbf{k}\cdot \mathbf{A}_{\pm,n}\mp1}.
\end{equation}

Inserting this into \eqref{Tij} and \eqref{Weinberg}, we can write the power radiated in the direction $ \mathbf{k}$ as a quadruple sum over the unit vectors $ \mathbf{A}_{\pm,n}$

\begin{equation}
\label{dPnotsummedyet}
\frac{dP}{d\mathbf{k}}=\frac{4G\mu^2}{\pi^3}\sum_{n,n',
 q,q'=0}^{N-1} \Lambda_{ij,lp}D_{+,n}^{(i}D_{-,n'}^{j)}D_{+,q}^{(l}D_{-,q'}^{p)}\left(\sum_{m=1}^{\infty}\frac{1}{m^2} \left(e^{im(-\alpha_n^+-\alpha_{n'}^-+\alpha_q^++\alpha_{q'}^-)}+ \text{similar terms}\right)\right).
\end{equation}
All $16$ terms in the parenthesis above come from the expansion of $(e^{-i m \alpha_n^+}-e^{-i m \widetilde{\alpha}_n^+})(e^{-i m \alpha_{n'}^-}-e^{-i m \widetilde{\alpha}_{n'}^-})(e^{i m \alpha_q^+}-e^{i m \widetilde{\alpha}_q^+})(e^{i m \alpha_{q'}^+}-e^{i m \widetilde{\alpha}_{q'}^-})$ so they are all of the form $\pm e^{im\gamma}$. Furthermore, $\frac{dP}{d\mathbf{k}}$ has to be real so we can replace all terms like $e^{im\gamma}$ by $\cos(m\gamma)$. We can now use the following summation formula to simplify \eqref{dPnotsummedyet}
\begin{equation}
\label{cossum}
\sum_{m=1}^{\infty}\frac{\cos(m\gamma)}{m^2}=\frac{1}{4}\left((\gamma~mod~2\pi)-\pi\right)^2-\frac{\pi^2}{12}.
\end{equation}
Since $8$ of the $16$ terms have a plus sign while the other $8$ have a minus sign in front, the contributions from $-\frac{\pi^2}{12}$ in \eqref{cossum} cancel in \eqref{dPnotsummedyet} and we are left with 
\begin{equation}
\label{dPsummed}
\frac{dP}{d\mathbf{k}}=\frac{G\mu^2}{\pi^3}\sum_{n,n',q,q'=0}^{N-1} \Lambda_{ij,lp}D_{+,n}^{(i}D_{-,n'}^{j)}D_{+,q}^{(l}D_{-,q'}^{p)} \Bigg[\bigg(\big((-\alpha_n^+-\alpha_{n'}^-+\alpha_q^++\alpha_{q'}^-)~mod~2\pi\big) -\pi\bigg)^2+ \text{similar terms}\Bigg]
\end{equation}

In the cases where the direction of emission is exactly equal to some $ \mathbf{A}_{+,n_0}$ (resp. to some $ \mathbf{A}_{-,q_0}$), this expression has to be corrected by removing all terms with $n=n_0$ or $n'=n_0$ (resp. $q=q_0$ or $q'=q_0$) in the quadruple sum. \\

The number of operations that one has to perform in order to compute numerically $\frac{dP}{d\mathbf{k}}$ for any given direction $\mathbf{k}$ grows as $\mathcal{O}(N^4)$ and quickly becomes prohibitive. The computation can be made faster by using some symmetries but $N$ of the order of $100$ seems a reasonable limit if one wishes to perform the integration in \eqref{totalpower} to obtain the total power radiated\footnote{As an alternative to our procedure, we could have used the results of \cite{Allen:1994iq} with the particular advantage of getting rid of the numerical integration, both a potential source of error and a time consuming step. We only chose to stick to our direct generalization of \cite{Garfinkle:1987yw} because it is simpler and because the smoothness of the power as a function of $\mathbf{k}$ enabled us to perform the integral with a reasonably small grid.}. 

\subsubsection{Results}
\begin{figure}[h]
\centering
   \includegraphics[scale=.7]{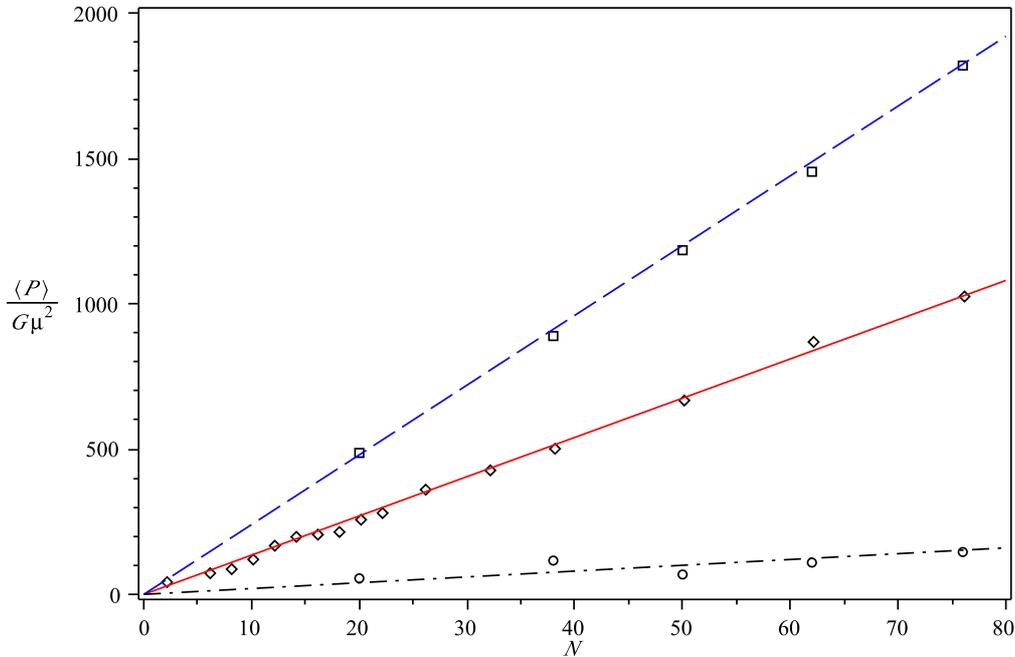} 
   \caption{Average power radiated by a loop with $N$ kinks in units of $G\mu^2$ as a function of $N$. The diamonds correspond to the distribution of loops defined in Sec. \ref{subsubsec:distrib} for which the distribution of kink angles is broadly peaked around $\pi/2$ and are very well fitted by the red straight line illustrating the linear growth at large $N$ found in Sec. \ref{subsec:analytical}. The top and the bottom set of points (and the associated dashed blue and dotdashed black straight lines) correspond to piecewise constant loops with fixed kink angles of respectively $135°$ and $\pi/10$, showing that the power radiated increases with the amplitude of the kinks.}
   \label{plotpowerofN}
\end{figure}

Combining the procedure described in Sec. \ref{subsubsec:distrib} to draw random loops in our class and the method of the previous paragraph to compute their power, we can now statistically estimate $\left \langle P \right \rangle$ for various values of $N$. The results, shown in diamonds in Fig.\ref{plotpowerofN} and well fitted by the solid (red) line, confirm the linear behavior derived in Sec. \ref{subsec:analytical} when the number of kinks becomes large. The best fit of our data with a linear function yields a value for the slope equal to $13.5$, showing good agreement with \eqref{PdominantNnum}.
We also observed that the ratio of the standard deviation to the average value decreases as $N$ grows so that the average value should be a good estimator of the typical power radiated.

The distribution defined in Sec. \ref{subsubsec:distrib} is flat in the sense that each of the $\mathbf{A}_{\pm,n}$ has a uniform probabilty on the unit sphere (or rigorously half of them, in order to obtain a closed loop). This induces a whole distribution of kink angles, broadly peaked around the typical value $\pi/2$. 

We would now like to know how the amplitude of the kinks affects the power radiated. We therefore modify the procedure to restrict ourselves to loops with fixed kink angles in the following way. Given $\alpha$ and two random unit vectors $\mathbf{A}_{\pm,0}$, we randomly draw each unit vector $\mathbf{A}_{\pm,n+1}$ (for $n$ between $0$ and $N/2-2$) with uniform probability \emph{among those that satisfy} $\mathbf{A}_{\pm,n}\cdot\mathbf{A}_{\pm,n+1}=\cos \alpha$. With the usual definitions in the literature (see for example \cite{Binetruy:2010bq}), this corresponds to a kink amplitude (or sharpness) $A=\frac{1}{2}||\mathbf{A}_{\pm,n}-\mathbf{A}_{\pm,n+1} ||=\sin(\alpha/2)$. Note that strictly speaking, not \emph{all} of the $2N$ kinks on these loops have an angle $\alpha$ but only $2N-4$, the other $4$ being potentially significantly different.

The results are shown in Fig.\ref{plotpowerofN} for two different values of $\alpha$, respectively $135°$ for the upper set of points (boxes) and $\pi/10$ for the lower set (circles)\footnote{The value $135°$ was chosen in order to compare our results with those of \cite{Quashnock:1990qy}. In addition to their argument based on the quadrupole radiation formula, the authors performed numerical simulations of loops with $8$, $16$ and $32$ kinks with opening angles of $135°$ to compute the numerical proportionality constant ($\Gamma_k$ in their notation). Note that these were simulations of the dynamics of Nambu-Goto loops including gravitational backreaction, an approach completely different from the numerical computations of the radiated power that we performed in this section. 
Taking into account the slightly different definition of the word kink used in their paper which introduces a factor of $2$, their results translate into a value $\Gamma_k\approx25-50$ for the slope corresponding to this particular value of the kink angle.} Here again, we find that the average power increases linearly with the number of kinks. We note however that the standard deviation becomes important at small $\alpha$ so for a given loop, the actual power radiated can be significantly larger than the average value.\footnote{This is due to the following: when $\alpha$ is small, the $\mathbf{A}_{+,n}$ will tend to be gathered together, densely populating a small region instead of covering the whole unit sphere. Now, in the cases where $\mathbf{A}_{+,0}$ and $-\mathbf{A}_{-,0}$ are drawn not too far from each other, these two regions intersect and the chances that one of the $\mathbf{A}_{+,n}$ becomes really close to one of the $-\mathbf{A}_{-,n}$ resulting in an extremely high instantaneous velocity and important radiation (an equality would result in a degenrate cusp as discussed in Sec. \ref{subsec:powerradiated}) will be high.} The best fits give slopes equal to $24$ for $\alpha=135°$ and $2$ for $\alpha=\pi/10$. As could be expected, we find that the power increases with the kink angle.

\subsection{Application to loops with junctions}
\label{subsubsec:complit}
We now conclude this section with a few remarks on the applicability of our results in the context of kink proliferation as described in \cite{Binetruy:2010cc} which occurs on non-periodic loops containing junctions and gives rise to a whole distribution of kink angles. First, we argue that although we have worked here only with closed periodic loops, the periodicity of the functions $\mathbf{x}_\pm$ does not play an important role and as a consequence, we expect our results to hold for the loops with junctions. Indeed, the fact that $\mathbf{x}_\pm$ are periodic induces correlations within each set of vectors $\mathbf{A}_{+,n}$ and $\mathbf{A}_{-,n}$ (in our class of loops, this translates into the choice $\mathbf{A}_{n,\pm}=-\mathbf{A}_{n-N/2,\pm}$) but these correlations only give rise to the second and the third term in \eqref{contractethreeterms} and those are subdominant at large $N$.

A more problematic issue is that of the value of $\Gamma'$ corresponding to the complicated distribution of kink angles produced during kink proliferation. The population of kinks resulting from proliferation can be divided into a number $k'$ of large amplitude kinks which dominate the signal received by an observer from a whole network of loops and a huge number of unobservable very small kinks. The tools that we developed in this section do not enable us to compute $\Gamma'$ for such an intricate distribution: because it is only known numerically, no analytical approach like that of Sec. \ref{subsec:analytical} is possible and the number of kinks exceeds by far the possibilities of the numerical approach of Sec. \ref{subsubsec:num}. However, we can still put a lower bound on the power radiated by such loops by considering only the large amplitude kinks. In this case, we have according to our results $P\sim G\mu^2 \Gamma'  k'$ with a value of $\Gamma'$ of the order of $10$. Given that we will only be interested in order of magnitude estimates in the remainder of this paper and that several parameters already suffer from comparable uncertainties, we can replace $\Gamma'$ in the expression of $P$ by the usual constant $\Gamma\sim50$ expressing the power radiated by smooth loops in units of $G\mu^2$ in order to simplify our expressions. This lower bound will be a reliable estimate of $P$ if the contribution from the small amplitude kinks is negligible. If instead the power is actually dominated by the small amplitude kinks, then $P$ will depend on their number rather than on $k'$. In this case, we can only conclude that $P\gg\Gamma G \mu^2k'$. Distinguishing between those two situations is not possible with the tools that we have developed so far and is a model-dependent issue since different types of strings may give rise to different kink amplitude distributions. As a consequence, in the rest of this paper, we will discuss the implications of both scenarios. The conclusions of this discussion can be summarized as follows: 
\begin{equation}
\label{Plargesmall}
\left\{\begin{array}{cc}P\sim \Gamma G\mu^2\cdot k' & \text{Power dominated by large kinks} \\P\gg \Gamma G\mu^2\cdot k' & \text{Power dominated by small kinks}\end{array}\right.
\end{equation}

\section{Consequences on the GW signatures from kink proliferation}
\label{sec:obspredictions}
The effect of kink proliferation on the GW signal emitted by a network of cosmic strings was computed in \cite{Binetruy:2010cc} in the general framework of a network where two types of loops coexist:
"standard" loops with a number of kinks of order unity, and loops with junctions\footnote{In this whole section, slightly abusing language, we take this expression to include also the periodic daughter loops (with no junctions) that may be formed due to the possible unzipping of junctions. In particular, when we talk about the lifetime of loops with junctions, we mean the time of decay by gravitational radiation of the loops formed with junctions and of their possible daughter loops obtained by junction unzipping and not the time of unzipping itself.} that have gone through a kink-proliferation phase. The average number of kinks per loop of the second subnetwork was taken to be a free parameter $k'\gg1$. 

Obviously, the predictions for the signal depend crucially on the number density of each type of loop since the rate of bursts is directly proportional to this quantity. In this section, we revisit the results of \cite{Binetruy:2010cc}, improving the estimation of the density of loops with $k'$ kinks by taking into account the fact that they radiate more gravitational power than initially considered and therefore decay faster. In Sec. \ref{subsec:densities}, we derive our new expression for the density of loops which is reduced at least by a factor $k'$. The consequences of this lower density on the gravitational wave bursts and on the stochastic background are studied in sections \ref{subsec:bursts} and \ref{subsec:background} respectively.

\subsection{Lifetime of loops with many kinks and density of loops}
\label{subsec:densities}

\subsubsection{One-scale model - standard case}

The predictions in \cite{Binetruy:2010cc} are based on a modified (in order to account for the presence of junctions) version of the standard one-scale model for the network evolution that was used for the original estimates \cite{Damour:2004kw,Damour:2001bk,Siemens:2006vk}. We start here by summarizing the key features of this model (see \cite{ShellardVilenkin} for a detailed discussion, together with \cite{Damour:2004kw} for for instance the factors of $p$).

A string network is made of both long strings and subhorizon closed loops. The long string component can be characterized by the typical distance between long strings $L^{st}$, the typical distance $\xi^{st}$ beyond which the directions of string segments become uncorrelated and the typical wavelength of the smallest wiggles $l_{\rm wiggles}$ on the long strings. In the so-called scaling regime, all these lengths are proportional to the Hubble radius and we have $L^{st}\sim p^{1/2} t$ and $\xi^{st}\sim t$. The value of $l_{\rm wiggles}$ is the subject of intense debate in the community and is usually written in terms of the parameter $\alpha$ under the form $l_{\rm wiggles}\sim\alpha t$.  Using a crude estimate of the efficiency of gravitational backreaction to round off the small scale structure on the long strings \cite{BouchetBennett} gives $\alpha\sim\Gamma G \mu$ where the constant is roughly equal to 50.\\

At any given time, loops are chopped off from (self-) intersection of long strings. Those large loops then fragment into many small loops (which give the dominant contribution to the GW signal) and whose typical size $l$ is that of the smallest wiggles on the long strings $l\sim\alpha t$. Under this assumption, the population of loops can be determined by remarking that the scaling regime requires a given rate of energy loss for the infinite string network ($p^{-1}\mu t$ per Hubble volume and per Hubble time) so the number of such loops formed in a Hubble time and within a Hubble volume has to be $p^{-1}\mu t/ (\mu l)\sim p^{-1}\alpha^{-1}$. These loops then decay by gravitational radiation. The power radiated by these smooth loops has been computed in \cite{Burden1985277,Vachaspati:1985,Garfinkle:1987yw} and is proportional to $G \mu^2$ with the same proportionality constant $\Gamma\approx50$ as in the previous paragraph. The lifetime of loops formed at time $t$ is therefore $\tau\sim\frac{\mu L}{\Gamma G \mu^2}\sim \frac{\mu \Gamma G \mu t }{\Gamma G \mu^2}\sim t$. In other words, all the loops formed at a given time $t$ decay in a Hubble time and with good approximation all loops have the same size $\alpha t$ at any time $t$ which leads to a distibution of loop lengths approximately equal to a delta function $n(l,t)\sim\frac{1}{p\alpha t^3}\delta(l-\alpha t)$ that we will write from now on $n(t)\sim\frac{1}{p\alpha t^3}$. \\

Some authors have computed the gravitational wave predictions in different scenarios for the evolution of the loop network, for instance including smaller loops with $\alpha=\epsilon \Gamma G \mu$ and $\epsilon\ll1$ or large loops \cite{Siemens:2006vk,Damour:2004kw,Damour:2001bk}. For the sake of simplicity we will however stick, as in \cite{Binetruy:2010cc}, to a modification of this simplest scenario to analyze the impact of kink proliferation.

\subsubsection{One-scale model - with proliferation: old version}
\label{subsubsec:oldonescale}
Several authors have studied the evolution of a network of cosmic strings with junctions, both analytically and numerically \cite{Tye:2005fn,Avgoustidis:2007aa,Avgoustidis:2009ke,Copeland:2005cy,Hindmarsh:2006qn,Urrestilla:2007yw,Rajantie:2007hp,Sakellariadou:2008fk} and found that the scaling behaviour of the infinite strings still exists in the presence of junctions. Based on these results, the one-scale model presented in the previous section was modified in \cite{Binetruy:2010cc} to take into account the population of loops with junctions in the case of a network containing different types of strings of comparable tensions and reconnection probabilities.
The simplifying picture adopted there in order to isolate the effect of kink proliferation was that despite the formation of junctions, all the quantities related to the infinite string network remained unchanged.

We are interested in the population of small loops and in particular in the fact that now two types of loops can form: standard periodic loops and loops with junctions on them. Because both types of loops are produced as in the standard case by iterative fragmentation of larger loops, it was argued that their typical size was that of the smallest wiggles on the infinite network $\alpha t$. In order to describe these two different types of loops, a new parameter $q$ defined as the fraction of loops formed with junctions was introduced.\\

As in the standard case, loop production is the mechanism by which the infinite string network loses energy and in order to ensure a scaling behavior, the total energy lost has to remain unchanged with respect to the standard case. Because both types of loops have the same energy, the total number of loops formed per Hubble time and within a Hubble volume remains $1/(p\alpha)$. However, this number now divides into $(1-q)/(p\alpha)$ standard loops and $q/(p\alpha)$ loops with junctions. Because the creation of a loop with junctions can only happen by fragmentation on a larger loop around a junction whereas the production of a standard loop can happen anywhere on the loop, we expect $q\ll1$ and the exact value of this parameters is probably model dependent since it is related to the tendancy of a particular type of string to create junctions. When working with order of magnitude estimates, $1-q$ can be replaced by $1$ in the number of standard loops. Furthermore, in the scaling regime, $q$ is expected to be time-independent.\\

We are interested in the number of kinks on each type of loop. The standard loops are the product of the fragmentation of loops without junctions and therefore, as in the standard case, should have a number of kinks of order unity. On the second type of loops, the presence of junctions leads to a quick proliferation of the number of kinks and we call $k'$ the average number of large amplitude kinks on these loops. Note that, as discussed in more detail in Section \ref{sec:prolpower}, because the proliferation of kinks is exponential, $k'$ is roughly equal to the number of kinks at the end of proliferation regardless of how the the duration of proliferation compares with the lifetime of the loop.\\

The last step to obtain the number density for each type of loop is to take into account the decay of these loops by gravitational radiation.  At that point, \cite{Binetruy:2010cc} made the conservative assumption that the power radiated by both types of loops was the same as in the standard scenario ($P=\Gamma G \mu^2$) and in particualar that the power radiated by the second type of loops was not affected by kink proliferation. As in the standard scenario, this leads to a number density 
\begin{equation}
n(t)\sim\frac{1}{p\alpha t^3}
\end{equation}
for standard loops and
\begin{equation}
n'_{\rm old}(t)\sim\frac{q}{p\alpha t^3}
\end{equation}
for loops with junctions.

\subsubsection{One-scale model - with proliferation: revisited estimates}
Since the loops with junctions are created with an invariant length $l\sim\alpha t\sim\Gamma G \mu t$, then their initial energy is $\mu \alpha t= \Gamma G \mu^2 t$. Combining this with our expression in \eqref{Plargesmall} for the power radiated by those loops, we find that for models where the power is dominated by the large kinks, the loops decay\footnote{This simple calculation implicitely assumes that the number of kinks remains constant throughout the evolution of the loop. Though this is not necessarily the case, the analysis of Sec. \ref{sec:prolpower} shows that because the number of kinks grows exponentially, our estimate for the lifetime of the loop is valid.} in a time
\begin{equation}
\tau_{l}\sim\frac{\Gamma G \mu^2 t}{\Gamma G \mu^2 k'}\sim\frac{t}{k'}.
\end{equation}
In the case where the small kinks provide the main contribution to the loss of energy from the loop, this value becomes an upper bound according to \eqref{Plargesmall}. In both cases, this is much shorter than the value $\tau_l=t$ that was used in \cite{Binetruy:2010cc}. Therefore, at any given time, only a fraction of all the loops considered in \cite{Binetruy:2010cc} will be present in the network and the number density of loops with many kinks is at most
\begin{equation}
n'(t)_{\rm new}\sim\frac{1}{k'}\frac{q}{p\alpha t^3}.
\end{equation}

We now analyze the effect of this smaller density on our results.

\subsection{Rates and individual bursts}
\label{subsec:bursts}
The goal of this subsection is not to compute in detail the rate of bursts but rather to explain qualitatively the consequences of the shorter lifetime on the predictions for the GW signal. Therefore, instead of explicitely describing the dependence of the rate of kink bursts on the various parameters and on redshift (see sections 3 and 4 of \cite{Binetruy:2010cc} for more detail), we will only be interested here in the intuitive fact that it is proportional to the density of loops and to the number of kinks per loop. Since all the other factors entering the rate are equal for both types of loops, this approach will be sufficient for our purpose of comparing the contribution from standard loops and from loops with junctions in our model.\\

Following the results of \ref{subsubsec:oldonescale}, it was found in \cite{Binetruy:2010cc} that
\begin{equation}
\left(\frac{\dot{N}_{\rm junctions}}{\dot{N}_{\rm standard}}\right)_{\rm old}\sim\frac{q\cdot k'\cdot p^{-1}\alpha^{-1}t^{-3}}{p^{-1}\alpha^{-1}t^{-3}}\sim qk'
\end{equation}
with the conclusion that for models satisfying the condition $qk'\gg1$, the main contribution to the rate of kink bursts came from the loops with junctions.\\

Now using our revisited one-scale model taking into account the faster decay of loops with many kinks, we find that this fraction is smaller by at least a factor $k'$
\begin{equation}
\left(\frac{\dot{N}_{\rm junctions}}{\dot{N}_{\rm standard}}\right)_{\rm new}<\frac{q\cdot k'\cdot p^{-1}\alpha^{-1}t^{-3}/(k')}{p^{-1}\alpha^{-1}t^{-3}}\sim q
\end{equation}
and because $q<1$, the contribution from standard loops \emph{always} dominates the rate of kink events, making immediately clear that the original predictions made by \cite{Damour:2004kw,Damour:2001bk}  concerning the observability of individual bursts apply.\\

Note that this leaves the final conclusions of \cite{Binetruy:2010cc} on the observability of individual kink bursts unchanged. Indeed, it was argued there that the usual procedure to compute the rates and amplitudes had to be revised by introducing a lower redshift cutoff $z_c$ below which no statistical approach was valid due to the lack of loops in this region\footnote{Also note, though this is not necessary here to conclude that the dominant kink signal comes from the standard loops, that the smaller redshift cutoff applying to loops with junctions is now larger than in \cite{Binetruy:2010cc} by a factor $(k')^{1/3}$ or larger.}. This led to the conclusion that the presence of kinky loops could only yield amplitudes for kinks bursts larger than the standard predictions in \cite{Damour:2001bk} in specific regions of parameter space where pulsar constraints already forbid this anyway.

\subsection{Stochastic background}
\label{subsec:background}
For any given type of burst, the rate of events coming from high redshifts is so large that an observer cannot distinguish the individual bursts and only sees their incoherent superposition as a stochastic background. Damour and Vilenkin \cite{Damour:2001bk} showed that the sum over the overlapping bursts could be written in terms of a redshift integral of the square amplitude of individual bursts emitted at redshift $z$ weighted by the number of bursts from that redshift. The crucial point is that one has to start the integral at the redshift $z_{b\rightarrow b}$ that corresponds to the minimal rate for the bursts to actually overlap at the detector. The integral ends at $z_{hf}$ beyond which the loops are so small that they only emit at frequencies larger than the frequency of observation. Obviously, this calculation is only valid for redshifts larger than $z_c$. This reads
\begin{equation}
\label{intbackground}
h_c^2(f)=\int_{{\rm max}(z_c,z_{b\rightarrow b})}^{z_{hf}}  h^2( z)  f^{-1}d\dot{N}(z).
\end{equation}

Our goal here is to investigate the modifications with respect to the results of \cite{Binetruy:2010cc} induced by the smaller density of loops with junctions $n'$. Three of the quantities appearing in \eqref{intbackground} depend on $n'$, namely $z_c$, $z_{b\rightarrow b}$ and $d\dot{N}(z)$ (the rate of bursts coming from the redshift interval $[z,z+dz]$). However, both in the case of the kink background and of the kink-kink background (superposition of bursts emitted at kink-kink encounters on the strings), the integral is always dominated by the largest redshift $z_{hf}$ (see Appendix B of \cite{Binetruy:2010cc}) which only depends on the size of the loops and therefore remains unchanged. As a result, the only corrections come from the factor $d\dot{N}(z)$ which is directly proportional to $n'$ in the case of loops with junctions and now smaller by a factor $k'$ or larger. This means that the final formulas for the amplitude of the kink and of the kink-kink background (respectively formulas (53) and (57) of \cite{Binetruy:2010cc}) need to be divided by at least a factor $\sqrt{k'}$.\\

Let us start by discussing the kink background. We now have, for the contribution from loops with junctions in the case where the large kinks dominate the power that they radiate,
\begin{equation}
h_c^{\rm kink}\sim q^{1/2} \cdot10 (G\mu) \alpha^{1/2} p^{-1/2} \Theta\left(f-\frac{1}{\alpha t_0}\right)\left(1+\alpha t_0 f\right)^{-4/3} \left(1+\frac{\alpha t_0 f}{z_{eq}^{3/2}}\right)^{1/3} \, ,
\label{hckinkinterpolating}
\end{equation}
the right hand side being an upper bound for $h_c^{\rm kink}$ in the case of a small kink dominated power. The contribution of the standard loops is obtained by substituting $q=1$ in the previous equation. Since $q<1$, we now find that the \emph{kink stochastic background is actually always dominated by the standard loops}, even in the models where the product $qk'$ is larger than $1$.\\

We now turn to the kink-kink background for which our conclusions will be different depending on whether the power radiated by the loops with junctions is dominated by the large kinks or by the small ones.

In the first case, the contribution from loops with junctions obtained in \cite{Binetruy:2010cc} has to be divided by $\sqrt{k'}$ and now reads
\begin{equation}
\label{hckk}
h_c^{\rm k-k}\sim  (qk')^{1/2}\cdot 10 (G\mu) \alpha^{1/2} p^{-1/2} \Theta\left(f-\frac{1}{\alpha t_0}\right)\left(1+\alpha t_0 f\right)^{-3/2} \left(1+\frac{\alpha t_0 f}{z_{eq}}\right)^{1/2} \, .
\end{equation}
Here, the contribution from standard loops is recovered by setting $q=k'=1$ and so, the conclusion that it is dominated by loops with junctions provided that $qk'>1$ remains valid. Furthermore, it is easy to check that in this case the k-k bacground remains larger than both the cusp and the kink backgrounds computed in the standard case. Indeed, we now have $h_{c}^{k-k \text{ (new)}}\sim h_{c}^{k-k \text{ (old)}}/ \sqrt{k'}$ and $h_{c}^{k \text{ (new)}}\sim h_{c}^{k \text{ (old)}}/\sqrt{qk'}$. On the other hand, it was shown in \cite{Binetruy:2010cc} that $h_{c}^{k-k \text{ (old)}}/h_{c}^{k \text{ (old)}}\sim \sqrt{k'}$. Combining these equations leads to $h_{c}^{k-k \text{ (new)}}/h_{c}^{k \text{ (new)}}\sim \sqrt{qk'}$.  Note that $h_{c}^{k \text{ (new)}}$ corresponds to the amplitude of the standard kink background, and that, according to \cite{Olmez:2010bi}, it is of the same order of magnitude as the standard cusp background.

The case where the small kinks dominate is more complex because the estimate in \eqref{hckk} now becomes an upper bound and it is impossible to compare the contributions from standard loops and loops with junctions as well as the backgrounds from different types of bursts. For now we simply notice that if the radiation from the small kinks is too large, then even in models with $qk'>1$, the kink-kink background will not be the dominant one.\\

From an observational point of view, the conclusions of this section are that the amplitude of the kink-kink background is reduced with respect to the results of \cite{Binetruy:2010cc} by a factor $\sqrt{k'}$ at least and that the region of parameter space for which the dominant background remains the kink-kink could be smaller than computed in \cite{Binetruy:2010cc}.

\section{Kink proliferation versus power radiated: towards estimating $k'$}
\label{sec:prolpower}
Our results on the power radiated by a loop with many kinks enables us to discuss the order of magnitude of the parameter $k'$ entering the observational predictions of Sec. \ref{sec:obspredictions}. 
\subsection{Unspoiled exponential kink proliferation}
Let us consider a loop with junctions of typical length $L$ undergoing kink proliferation due to the transmissions and reflexions of kinks propagating through the junctions. We neglect for now all the potential mechanisms identified in \cite{Binetruy:2010bq} that could alter the course of proliferation, namely the self-intersections of the loop which could chopp off smaller loops without junctions but with many kinks, the unzipping of junctions and the gravitational backreaction (by this we mean the deviations to the Nambu Goto dynamics on a fixed flat background and in particular the rounding off of the kinks; as explained below, we however take into account the fact that the energy of the loop is indeed decreasing). We discuss the effect of these additional physical ingredients in Sec. \ref{subsec:othereffects}. Under these assumptions, it was shown in \cite{Binetruy:2010bq} that the number of sharp kinks $N(t)$ on the loop grows exponentially:
\begin{equation}
\label{numberkinksprolif}
N(t)\sim\ \rho ^{t/L}
\end{equation}
where the origin of time was set at the formation of the loop. The coefficient $\rho>1$ depends on the tensions of the strings in the loop. In the case where all the strings have the same tension, $\rho$ is maximal and roughly equal to $2$.\\

As the number of kinks increases on the loop, it radiates more power until, at $t=\tau_f$ it has lost all its initial energy which is roughly equal to $\mu L$. 

We start by discussing the case where the power radiated by the loop is dominated by the large amplitude kinks \emph{during the whole course of proliferation}. Using \eqref{Plargesmall}, we find that between $t$ and $t+ dt$, the loop looses an energy $dE\sim \Gamma G \mu^2\cdot N(t) dt$. Therefore, $\tau_f$ satisfies the equation

\begin{equation}
\Gamma G \mu^2\cdot \int_0^{\tau_f} N(t) dt\sim\mu L
\end{equation}
which, combined with \eqref{numberkinksprolif}  leads to
\begin{equation}
\label{Ntauf}
N(\tau_f)\sim  \frac{\ln \rho}{\Gamma G \mu} \sim  \frac{1}{\Gamma G \mu}.
\end{equation}
provided $\rho$ is not too close to $1$.\\

How is this related to the parameter $k'$ appearing in the computations of the gravitational wave signal in Sections \ref{subsec:bursts} and \ref{subsec:background} for instance? The number of kinks does not remain constant throughout the lifetime of the loop and $k'$ has to be defined as the average number of kinks per loop which is equal to the time average of the number of kinks on a given loop. Therefore, $k'$ will be somewhat smaller than $N(\tau_f)$ but, given the exponential growth of the number of kinks, we expect it to be comparable. This gives us our final estimate for $k'$

\begin{equation}
\label{kprimemax2}
k'\sim ( \Gamma G \mu)^{-1}.
\end{equation}

In the case where the small kinks actually provide the main contribution to the loss of energy from the loop \emph{for at least part of the proliferation phase}, our estimate turns into an upper bound

This is the number of kinks one obtains by assuming that proliferation occurs in the ideal conditions mentioned earlier. In other words, when the only limitation to proliferation is the decay of the loop by gravitational radiation. In this case, the duration of proliferation is equal to the lifetime of the loop. In \cite{Binetruy:2010bq}, various mechanisms which might be responsible for an alteration of the proliferation of kinks were proposed. We now discuss qualitatively how these can affect our estimate of $k'$.

\subsection{Discussion of other effects}
\label{subsec:othereffects}
\subsubsection{Unzipping of junctions}
The first mechanism is related to the potential unzipping of the junctions. In particular, simulations of realistic loops with junctions in \cite{Binetruy:2010bq} showed that one of the segements of string typically ends up shrinking to a point and at that time, both junctions collide. What happens at that point in not encoded in the effective modified Nambu action that was used and certainly depends on the underlying theory \cite{Firouzjahi:2009nt,Bevis:2009az} but one possibility is that the junctions disappear or unzip to release two periodic closed loops. In that case, proliferation brutally stops and the number of kinks then remains constant. It is interesting to remark the typical time for this collision to happen (a few $L$ according to our simulations) is comparable with (and seems to be slightly smaller than) the lifetime predicted by considering the gravitational decay which, as can easily be seen from \eqref{numberkinksprolif}, is given by $\ln\ (\Gamma G \mu)^{-1}$ in units of $L$. Therefore, the number of kinks can be significantly smaller than that predicted in \eqref{kprimemax2}.\\

\subsubsection{Self-intersections}
Another physical phenomenon that we neglected is the possibility for the segments of string to self-intersect during the evolution and reconnect to chopp off small loops carrying many kinks which therefore do not enter the proliferation anymore. This effect has not been studied quantitatively for loops with junctions.
In particular, it is not clear to what extent the frequency of these self intersections will increase as the number of kinks grows. Moreover, the effect will depend on the value of the reconnection probability since a low value of $p$ would suppress the production of small loops even when many self intersections take place. In any case, as we explain below, we do not expect that multiple reconnections affect drastically the estimate for $k'$. Indeed, when a loop without junctions is chopped off, it withdraws kinks from the loop where proliferation takes place but also instantaneously reduces its length. This has two effects. It increases the rate of proliferation as the exponent in \eqref{numberkinksprolif} is now larger but it also decreases the energy of the loop and therefore its lifetime and the duration of proliferation.

Let us consider a very simplistic toy model to measure the importance of these competing effects. We start with a loop containing three segments of length $L$ meeting at two junctions. Suppose that at some instant $t_0$, each of the three segments self-intersects and has its length reduced by a factor $\alpha<1$ (the length of the loop released is then $(1-\alpha)L$). Assuming that the kinks are evenly spread over the segment, the number of kinks on each segment right after reconnection is $N_0=\alpha\ \rho^{t_0/L}$ and from then on, it evolves as $N(t)=N_0 \rho^{(t-t_0)/(\alpha L)}$. Solving for the number of kinks at time $\tau_f$ when the loop has lost all its energy by gravitational radiation, we recover the same result as in the case with no reconnection $N(\tau_f)\approx(\Gamma G \mu)^{-1}+N_0\approx(\Gamma G \mu)^{-1}$. The value of $k'$ is slightly increased by the presence of kinks in the three loops produced at the reconnections but as there is no proliferation on these, we expect this contribution to be subdominant.\\

\subsubsection{Backreaction}
While we accounted for the loss of energy by gravitational radiation in our analysis, we completely neglected gravitational backreaction, namely the fact that this loss of energy causes the dynamics to deviate from the Nambu action on a fixed background. In our context, backreaction has at least two important effects. The first one is to round off the kinks \cite{Quashnock:1990wv,Hindmarsh:1990xi}, with a timescale that is probably of the order of the lifetime of the loop $\tau_f$ that we computed by looking at radiation. This tends to reduce the number of kinks in the system but at the same time, it increases the lifetime of the loop and therefore the duration of proliferation. We find however unlikely that this will contribute to increase the number of kinks with respect to \eqref{kprimemax2}. The second effect is that the string shrinks as it loses its energy, or in other words, its length is actually a (decreasing) function of time. This will certainly alter the $\rho^{t/L}$ behavior that was derived in \cite{Binetruy:2010bq} where $L$ was the initial length of the segments of string, in particular towards the end of the evolution where the rate of proliferation will become increasingly large.

\section{Conclusion}

In this paper, we have investigated the loss of energy by gravitational radiation from cosmic string loops with many kinks. Using analytical tools, we found that the power radiated by piecewise linear loops with $N$ kinks and a distribution of kink angles roughly peaked around the value $\pi/2$ is proportional to $N$ with a proportionality constant of the form $\Gamma' G \mu^2$ and the order of magnitude $\Gamma'\sim10$. We also investigated how our result depends on the typical amplitude of the kinks by computing numerically the power radiated by loops with fixed kink angles. We showed that the linear behavior in the number of kinks is robust and that $\Gamma'$ is an increasing function of the typical kink angle. This is the first time that radiation from loops with many sharp kinks was studied using a full relativistic formalism.

We then applied our results in the context of kink proliferation on loops with junctions, a phenomenon described in \cite{Binetruy:2010bq}. The distribution of kinks on loops with junctions is complex and hard to study even numerically, in particular because it contains a huge a number of very small kinks. For this reason, precisely estimating the power radiated by such loops is impossible. However, our results from Sec. \ref{sec:radiation} enabled us to determine a lower bound on this power by considering only the interaction of the $k'$ sharp kinks that dominate the contribution to the gravitational wave burst signal received by an observer from a whole cosmological network of loops. More precisely, we found that for loops with junctions, $P>\Gamma G \mu^2 \cdot k'$ where $\Gamma\sim50$ is the standard value of the power radiated by smooth loops in units of $G\mu^2$. For models where the contribution from the small kinks to the power radiated is subdominant, this lower bound will be saturated. In any case, the power radiated by loops with junctions is much larger than the estimate $P\sim \Gamma G \mu^2$ used in \cite{Binetruy:2010cc}. 

As explained in Sec. \ref{sec:obspredictions}, this affects the observational predictions for the gravitational waves emitted by a network of cosmic string loops with junctions: since they radiate more, those loops decay faster and so their number density is smaller thereby decreasing the rates of events. We found that the rate of individual kink bursts should not be larger in networks with junctions than in standard networks. This strengthens the conclusion of \cite{Binetruy:2010cc} that, contrary to what one might expect at first sight, the presence of loops with junctions in a network and the associated kink proliferation do not provide any distinctive signature as far as the observation of individual bursts is concerned.

According to \cite{Binetruy:2010cc}, the main observational consequence of kink proliferation is the existence of a strong stochastic background of gravitational waves resulting from the superposition of the large number of bursts produced at kink-kink encounters. More precisely, it was shown that for models where the fraction of loops with junctions $q$ and the number of sharp kinks on them $k'$ satisfy the condition $qk'\gg1$, this kink-kink background dominates over the cusp and the kink one. Our analysis slightly weakens this conclusion: the characteristic amplitude of the kink-kink background is suppressed by at least a factor $\sqrt{k'}$ and so the region of parameter space in which it remains dominant may be reduced. We note however that for models where the total radiated power from the loops with junctions is dominated by the sharp kinks, the condition for the kink-kink background to dominate remains $qk'\gg1$.

Finally, our result enabled us to put an upper bound to the parameter $k'$ resulting from kink proliferation by computing the decay of a loop by gravitational radiation as the number of kinks grows. By assuming that exponential proliferation lasts until the loop has radiated all its initial energy away, we found that $k'<(\Gamma G \mu)^{-1}$. We also discussed qualitatively how other effects could affect this upper bound but ultimately,  a quantitative analysis combining all these effects would be necessary to provide a complete picture of the physics of kink proliferation and to estimate $k'$, a crucial value for the observational predictions. 

\section*{Acknowledgements}
We would like to thank Pierre Bin\'etruy, Thomas Hertog and Daniele Steer for many useful conversations during the completion of this work.

\appendix
\section{Expression of $\left \langle I_+^iI_+^{j*}  \right \rangle$ in terms of $s, \mu$ and $\sigma$}
\label{sec:appendixA}
In this appendix, we detail the calculation leading to Eq. \eqref{IIQ}. Our starting point is formula \eqref{I+}
\begin{equation}
I_+^i=\frac{i}{m}\sum_{n=0}^{N-1} C_n^i e^{\frac{2i\pi m}{N}\left(n-\mathbf{k}\cdot \left(\sum_{s=0}^{n-1} \mathbf{A}_s\right)\right)}
\end{equation}
where we recall that
\begin{equation}
\label{Cn}
C_n^i=\frac{A_n^i}{\mathbf{k}\cdot \mathbf{A}_n-1}\left(e^{\frac{2i\pi m}{N}(1-\mathbf{k}\cdot \mathbf{A}_n)}-1\right).
\end{equation}
This yields
\begin{equation}
\label{Idoublesum}
\left \langle I_+^iI_+^{j*} \right \rangle= \frac{1}{m^2}\sum_{n=0}^{N-1}\sum_{n'=0}^{N-1} S_{n,n'}^{i,j}
\end{equation}
with the definition
\begin{equation}
S_{n,n'}^{i,j}=\left \langle C_n^{i}C_{n'}^{j*} e^{-\frac{2i\pi m}{N}\mathbf{k}\cdot \left(\sum_{s=0}^{n-1} \mathbf{A}_s-\sum_{s=0}^{n'-1} \mathbf{A}_s\right)}\right \rangle e^{\frac{2i\pi m}{N}(n-n')}.
\end{equation}
Depending on the values of $n$ and $n'$, some of (or all) the vectors appearing inside the average brackets are independent so that the average conveniently factorizes. In order to make the computation simpler, and to reduce the number of different situations that we need to distinguish, we will use the symmetry property $S_{n',n}^{i,j}=\left(S_{n,n'}^{j,i}\right)^*$ to reduce the sum to terms with $n\leq n'$. This leads to

\begin{equation}
\label{Idoublesumsimplified}
m^2\left \langle I_+^iI_+^{j*} \right \rangle=\sum_{n=0}^{N-1} S_{n,n}^{i,j} + \sum_{0\leqslant n< n'\leqslant N-1} S_{n,n'}^{i,j}+ \left(S_{n,n'}^{j,i}\right)^*
\end{equation}

The terms in the first sum simply reduce to
\begin{equation}
S_{n,n}^{i,j}=\left \langle C_n^i C_{n}^{j*}\right \rangle.
\end{equation}

The second sum only contains terms with $n'> n$, which we can rewrite
\begin{equation}
S_{n,n'>n}^{i,j}=\left \langle C_n^{i}C_{n'}^{j*}\prod_{s=n}^{n'-1} e^{\frac{2i\pi m}{N}\mathbf{k}\cdot\mathbf{A}_s}\right \rangle e^{\frac{2i\pi m}{N}(n-n')}.
\end{equation}
 
We need to distinguish the following cases:\\
\begin{itemize}
\item $n'=n+N/2$: There are $N/2$ such terms. In this case, $ \mathbf{A}_n=- \mathbf{A}_{n'}$. All the factors between the brackets are uncorrelated except for $C_n^{i}$, $C_{n'}^{j*}$ and $e^{\frac{2i\pi m}{N}\mathbf{k}\mathbf{A}_n}$.
We then have 
\begin{equation}
S_{n,n+N/2}^{i,j}=(-1)^m\left \langle C_n^{i}C_{n+N/2}^{j*} e^{\frac{2i\pi m}{N}\mathbf{k}\cdot\mathbf{A}_n} \right \rangle \prod_{s=n+1}^{n-1+N/2} \left \langle e^{\frac{2i\pi m}{N}\mathbf{k}\cdot\mathbf{A}_s}\right \rangle
\end{equation}

\item $n'<n+N/2$: All the factors between the brackets are uncorrelated except for $C_n^{i}$ and $e^{\frac{2i\pi m}{N}\mathbf{k}\mathbf{A}_n}$.
In this case,
\begin{equation}
S_{n,n'<n+N/2}^{i,j}=\left \langle C_n^{i}e^{\frac{2i\pi m}{N}\mathbf{k}\cdot\mathbf{A}_n} \right \rangle \left \langle C_{n'}^{j*}\right \rangle \prod_{s=n+1}^{n'-1} \left \langle e^{\frac{2i\pi m}{N}\mathbf{k}\cdot\mathbf{A}_s}\right \rangle e^{\frac{2i\pi m}{N}(n-n')}
\end{equation}

\item $n'>n+N/2$: Some of the terms under the product sign cancel out and all the factors that are left are uncorrelated.
In this case,
\begin{equation}
S_{n,n'>n+N/2}^{i,j}=\left \langle C_n^{i} \right \rangle \left \langle C_{n'}^{j*}\right \rangle \prod_{s=n+1}^{n-n'+N} \left \langle e^{\frac{2i\pi m}{N}\mathbf{k}\cdot\mathbf{A}_s}\right \rangle e^{\frac{2i\pi m}{N}(n-n')}
\end{equation}

\end{itemize}

This leads us to define the following average values involving only one of the random vectors $\mathbf{A}_n$:
\begin{equation}
\label{sdef}
s=\left \langle e^{\frac{2i\pi m}{N}\mathbf{k}\cdot\mathbf{A}_n}\right \rangle
\end{equation}
\begin{eqnarray}
\label{rho}
\label{rhodef}
\rho^i & = &\left \langle C_n^i \right \rangle\\
\label{rhotildedef}
\tilde{\rho}^i & = & \left \langle C_n^i e^{\frac{2i\pi m}{N}\mathbf{k}\cdot\mathbf{A}_n}\right \rangle
\end{eqnarray}
\begin{eqnarray}
\label{sigma}
\sigma^{ij} &= &\left \langle C_n^i C_{n}^{j*}\right \rangle\\
\label{mu}
\mu^{ij}&=&\left \langle C_n^{i}C_{n+N/2}^{j*} e^{\frac{2i\pi m}{N}\mathbf{k}\cdot\mathbf{A}_n} \right \rangle
\end{eqnarray}
Given our procedure to draw the vectors $ \mathbf{A}_n$, it is clear that none of the average values defined here actually depends on the index $n$. Note however that all of them depend on $ \mathbf{k}$ and $m$ but we keep this dependence implicit here.

The expression of $ \left \langle I_+^iI_+^{j*} \right \rangle$ in terms of $\mu^{ij}$, $\sigma^{ij}$, $\tilde{\rho}^i$, $\rho^i$ and $s$ reads
\begin{equation}
\label{IIavecK}
m^2\left \langle I_+^iI_+^{j*} \right \rangle = N \sigma^{ij} + \frac{(-1)^m}{2}N \left( \mu^{ij} s^{N/2-1}+\mu^{ij*}(s^*)^{N/2-1}\right)+K_1 \rho^i \rho^{j*}+K_2 \tilde{\rho}^i \rho^{j*}+K_2^*  \rho^{i}\tilde{\rho}^{j*}
\end{equation}
with
\begin{eqnarray}
K_1&=&\sum_{n=0}^{N-1}\sum_{n'=n+N/2+1}^{N-1} e^{\frac{2i\pi m}{N}(n-n')} s^{N-n'}\\
K_2&=&\sum_{n=0}^{N-1}\sum_{n'=n+1}^{n+N/2-1} e^{\frac{2i\pi m}{N}(n-n')} s^{n'-n}\
\end{eqnarray}

In principle, we now need to compute the full tensors $\mu^{ij}$, $\sigma^{ij}$, $\tilde{\rho}^i$, $\rho^i$ as well as the value of $s$, $K_1$ and $K_2$.
However the number of components that we will really need can be drastically reduced by using the symmetries of our problem. As pointed out earlier, since there is no preferred direction in our class of loops,  $\langle\frac{dP}{d \mathbf{k}}\rangle$ will be independent of $ \mathbf{k}$ and we had set $\mathbf{k}=(1,0,0)$. Now remember that we will eventually contract $ \left \langle I_+^iI_+^{j*} \right \rangle$ with the projection operator $\Lambda_{i,j,l,p}$ which projects away any term proportional to $\mathbf{k}$. Therefore, any component of $\left \langle I_+^iI_+^{j*} \right \rangle$ bearing an index $1$ will be projected away. This means that we only need to calculate $\tilde{\rho}^2$, $\tilde{\rho}^3$,$\rho^2$, $\rho^3$, $\sigma^{22}$, $\sigma^{33}$, $\sigma^{23}$, $\sigma^{32}$, $\mu^{22}$, $\mu^{33}$, $\mu^{23}$ and  $\mu^{32}$. We can even go a little further by remarking that $\tilde{\rho}^i$ and $\rho^i$ can only be proportional to $k^i$ (the only vector in the game) so that 

\begin{equation}
\label{rhonuls}
\rho^2=\rho^3=\tilde{\rho}^2=\tilde{\rho}^3=0
\end{equation}
and the last three terms in \eqref{IIavecK} will be projected away. This means that we don't need to compute $K_1$ and $K_2$. Furthermore, directions $2$ and $3$ are equivalent so we have
\begin{equation}
\label{sigmamu}
\sigma^{22}=\sigma^{33}\equiv\sigma, \qquad \mu^{22}\equiv\mu^{33}=\mu,
\end{equation}
\begin{equation}
\label{sigmamunuls}
\sigma^{23}=\sigma^{32}=\mu^{23}=\mu^{32}=0.
\end{equation}
which only leaves us with the three quantities $\sigma$ and $\mu$ and $s$ to compute. For the sake of clarity, we postpone their computations to Appendix \ref{appendix:smusigmaasfmN}.

We can finally rewrite \eqref{IIavecK} in terms of $s, \mu$ and $\sigma$ to recover Eq. \eqref{IIQ}
\begin{equation}
\left \langle I_+^iI_+^{j*} \right \rangle=  \frac{N}{m^2} \left(\sigma+(-1)^m s^{N/2-1} \mu \right) \delta^{ij}(1-\delta^{i1})+ Q^{ij}
\end{equation}
where the tensor $Q^{ij}$ will be projected away by $\Lambda_{ijlp}$.

\section{Computation of $s$, $\mu$ and $\sigma$ as functions of $m$ and $N$}
\label{appendix:smusigmaasfmN}
In this appendix, we detail the calculation of the functions of $s$, $\mu$ and $\sigma$ defined in \eqref{sdef} and \eqref{sigmamu}. Each of them is an average value of a quantity that depends on only one of the unit vectors drawn randomly on the unit sphere. We parametrize this unit vector using spherical coordinates $(\theta,\phi)$, so that the vector components are $(\cos(\theta),\sin(\theta)\cos(\phi),\sin(\theta)\sin(\phi))$. Since we assumed a uniform distribution over the unit sphere, $\phi$ has to be uniformly distributed in the interval $[0,2\pi]$ while $\cos\theta$ is uniformly distributed in the interval $[-1,1]$. The average values now rewrite as double integrals over $\theta$ and $\phi$. 

Let us start with $\sigma$
\begin{eqnarray}
\sigma = \left \langle C_n^2C_{n}^{2*} \right \rangle & = & \left \langle \frac{(A_n^2)^2}{(\mathbf{k}\cdot \mathbf{A}_n-1)^2}\left|\left(e^{\frac{2i\pi m}{N}(1-\mathbf{k}\cdot \mathbf{A}_n)}-1\right)\right|^2 \right \rangle\\
 & = &4 \left \langle \frac{(A_n^2)^2}{(\mathbf{k}\cdot \mathbf{A}_n-1)^2} \sin^2\left(\frac{\pi m}{N}(1-\mathbf{k}\cdot \mathbf{A}_n) \right) \right \rangle\\
 & = & \frac{4}{4\pi}\int_{0}^{2\pi} \cos^2\phi~d\phi \int_{0}^{\pi} \sin\theta \frac{\sin^2\theta}{(1-\cos\theta)^2}\sin^2\left(\frac{\pi m}{N}(1-\cos\theta)\right)d\theta \\
 & = & \Cin\left(\frac{4\pi m}{N}\right) +\frac{N}{4\pi m}\sin\left(\frac{4\pi m}{N}\right) -1
\end{eqnarray}
where we have used the variable change $u=\frac{\pi m }{N}(1-\cos\theta)$ to compute the integral over $\theta$. In the last line, we have used the special cosine integral defined as
\begin{equation}
\Cin(x)=\int_0^x\frac{1-\cos t}{t}dt.
\end{equation}\\

In the same fashion, we can compute
\begin{eqnarray}
\mu&=&\left \langle C_n^{2}C_{n+N/2}^{2*} e^{\frac{2i\pi m}{N}\mathbf{k}\cdot\mathbf{A}_n} \right \rangle\\
& = & \left \langle \frac{(A_n^2)^2}{(\mathbf{k}\cdot \mathbf{A}_n-1)(\mathbf{k}\cdot \mathbf{A}_n+1)}\left(e^{\frac{2i\pi m}{N}(1-\mathbf{k}\cdot \mathbf{A}_n)}-1\right) \left(e^{-\frac{2i\pi m}{N}(1+\mathbf{k}\cdot \mathbf{A}_n)}-1\right) e^{\frac{2i\pi m}{N}\mathbf{k}\cdot\mathbf{A}_n}\right \rangle\\
  & = & \frac{4}{4\pi}\int_{0}^{2\pi} \cos^2\phi~d\phi \int_{0}^{\pi} \sin\theta \frac{\sin^2\theta}{(\cos\theta-1)(\cos\theta+1)}\sin\left(\frac{\pi m}{N}(1-\cos\theta)\right)\sin\left(\frac{\pi m}{N}(1+\cos\theta)\right)d\theta\\
& = & \cos\left(\frac{2\pi m}{N}\right)-\frac{N}{2\pi m}\sin\left(\frac{2\pi m}{N}\right).
\end{eqnarray}\\

Finally, we have
\begin{equation}
s=\left \langle e^{\frac{2i\pi m}{N}\mathbf{k}\cdot\mathbf{A}_n}\right \rangle=\frac{1}{4\pi}\int_{0}^{2\pi} d\phi \int_{0}^{\pi} \sin\theta e^{\frac{2i\pi m}{N}\cos\theta }d\theta=\frac{N}{2\pi m} \sin\left(\frac{2\pi m}{N}\right)
\end{equation}

Of course, one can check \eqref{rhonuls}, \eqref{sigmamu} and \eqref{sigmamunuls} using this procedure and variable changes adapted to the symmetry.

\bibliographystyle{utphys}
\bibliography{bibliofinal.bib}

\end{document}